# Optics in a field of gravity

E. Eriksen[#] and Ø. Grøn[#¤]

[#] Institute of Physics, University of Oslo, P.O.Box 1048 Blindern, N-0316 Oslo, Norway

[¤]Department of Engineering, Oslo University College, St.Olavs Pl. 4, 0130 Oslo, Norway.

E-mail: oyvind.gron@iu.hio.no

**Abstract**

An observer at rest in a uniformly accelerated reference frame experiences a parallel gravitational field, where the rays of light are formed as circular arcs instead of being straight lines. We deduce how a sphere at rest in the Rindler frame, and a freely falling sphere, will appear to an observer at rest in the frame.

## 1. Introduction

According to the principle of equivalence the physics in a parallel gravitational field is equivalent to the physics in a linearly accelerated reference frame in flat spacetime. Hence, in this article we shall present a simple investigation of the optics in a field of gravity by considering some optical phenomena in the Rindler frame, i.e. in a uniformly accelerated reference frame.

Surprisingly, this topic does not seem to have been investigated before. A search in Google of "optics in a field of gravity" gave no hits.

However an article appeared recently which inspired us to perform the present work. R. Beig and J. M. Heinzle[1] investigated the appearance of a sphere at rest in an inertial frame as observed by a uniformly accelerated observer. The investigation was performed with reference to the inertial frame by taking into account the effect of aberration.

In the present work we describe similar situations with reference to the Rindler frame in which the accelerated observer is at rest.

## 2. The Rindler frame

We consider a uniformly accelerated reference frame i.e. a Rindler frame R in flat spacetime where the reference particles have a constant proper acceleration. The Rindler co-moving co-ordinates $t, x, y, z$ are defined by the following transformation relative to the co-ordinates $T, X, Y, Z$ in an inertial frame, IF, (using units so that $c = 1$),

$$T = x \sinh gt \ , \quad X = x \cosh gt \ , \quad Y = y \ , \quad Z = z \tag{1}$$

where $g$ is constant, $x > 0$ and $-\infty < t < \infty$. The line element is

$$ds^2 = -g^2 x^2 dt^2 + dx^2 + dy^2 + dz^2 . \tag{2}$$

From eq.(1) follows that

$$X^2 - T^2 = x^2 . \tag{3}$$

Hence, the world lines of the reference particles in R, $x = $ *constant*, are hyperbolae.

Hyperbolic motion is characterized by a constant proper acceleration, i.e. constant acceleration as measured in an instantaneous rest frame. The reference frame moves in a Born-rigid way, meaning that the proper distance between the reference particles remain constant. Hence as measured in IF the distances between the reference particles of AF are Lorentz contracted. This means that the particles move faster the further behind they are, and there is a limit to the backwards extension of the Rindler frame defined by the condition that the reference particle at this limit moves with the velocity of light. The Rindler coordinates are chosen such that this limit appears at $x = 0$.

Keeping $x$ = constant in eq.(1) we get for the velocity and acceleration of a reference particle in R, as measured in IF,

$$v = \frac{dX}{dT} = \tanh gt \ , \quad a = \frac{dv}{dT} = \frac{1}{x} \frac{1}{\cosh^3 gt} \quad (4)$$

The corresponding components of the four-velocity and four-acceleration are

$$v^\mu = \frac{dX^\mu}{d\tau} = (\cosh gt, \ \sinh gt, \ 0, \ 0) \ ,$$
$$a^\mu = \frac{dv^\mu}{d\tau} = \frac{1}{x}(\sinh gt, \ \cosh gt, \ 0, \ 0) \ . \quad (5)$$

Thus,

$$a_\mu a^\mu = \frac{1}{x^2}(\cosh^2 gt - \sinh^2 gt) = \frac{1}{x^2} \quad (6)$$

Which gives a proper acceleration

$$\hat{a} = 1/x \quad (7)$$

or $c^2/x$ if we include the velocity of light. Hence, a fixed point in R has a proper acceleration $1/x$. This means that an observer at rest in R experiences an acceleration of gravity

$$\vec{a} = -\hat{a}\vec{e}_x = -(1/x)\vec{e}_x. \quad (8)$$

Note that $\hat{a} \to \infty$ when $x \to 0$. As observed in IF the reference particles with $x = 0$ according to eq.(3) approaches the origin with the velocity of light. Then at the point of time $T = 0$ they get an infinitely great acceleration making them move in the positive $X$-direction with the velocity of light. These particles, defining the limit of R, move along the $X$-axis like light being reflected by a mirror at $X = 0$. No particle can arrive from IF behind this limit and enter R. Hence, the surface $x = 0$ represents a horizon in R. The acceleration of gravity experienced in R is infinitely great at the horizon.

In the positive $X$-direction R extends infinitely far, and in the limit $x \to \infty$ the reference particles of R are permanently at rest in IF. The acceleration of gravity experienced in R vanishes in this limit.

## 3. Rays of light in the Rindler frame

The path of a photon in R is found most simply from the transformation (1). In IF the path is a straight line. We have shown in[2] that the orbit of a photon in R is a circular arc in a vertical plane with centre at the horizon, $x = 0$.

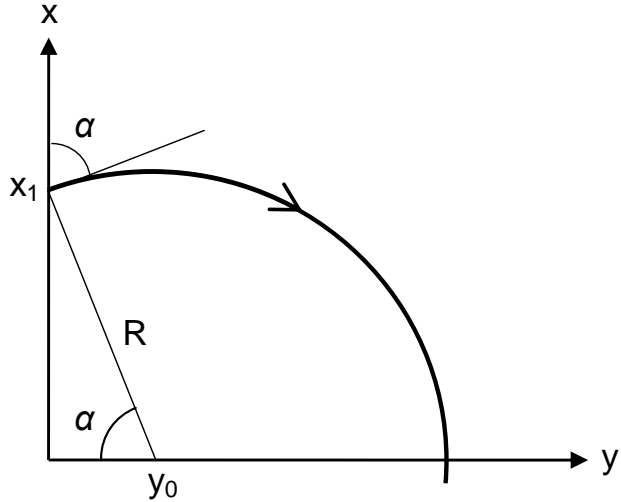

**Figure 1.** The figure shows a ray of light, or the path of a photon, emitted in the $xy$-plane from the point $x_1, 0, 0$ in a direction forming the angle $\alpha$ with the $x$-axis. Here $R = x_1 / \sin\alpha$, $y_0 = x_1 \cot\alpha$. If, for example, $\hat{a} = 10 m/s^2$, then $x_1 = c^2/\hat{a} = 0.95$ light years. Note that the path of an arbitrary photon hits the horizon plane of the Rindler space in an orthogonal direction.

## 4. Visual appearance of a sphere at rest in a field of gravity

We shall here find the appearance of a spherical source of light at rest in R as photographed from above and below.

### 4a. A sphere seen from above

The observer, P, is at a height $x = x_1$. The centre of the sphere is at a distance $b$ vertically beneath P. The radius of the sphere is $r$. Figure 2 shows the situation in an arbitrarily chosen vertical plane through the centre of the sphere. The circular arc is a light ray emitted tangentially from the surface of the sphere and passing through the observer, making an angle $\theta$ with the vertical as it arrives at P. The angle $\theta$ is the half of the apex angle.

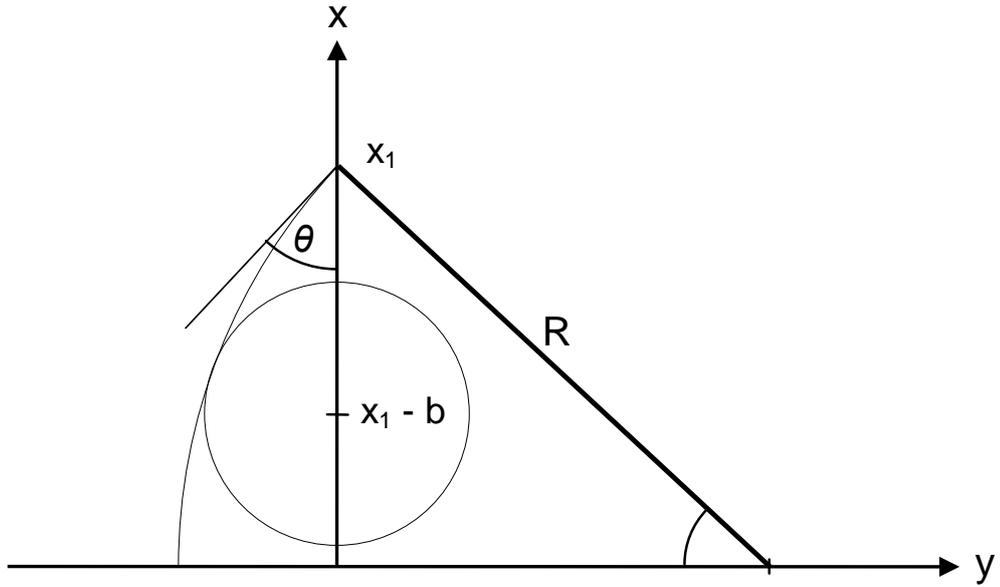

**Figure 2.** The figure shows the situation in an arbitrarily chosen vertical plane through the centre of the sphere. The circular arc is a light ray emitted from the sphere and passing through the observer, making an angle $\theta$ with the vertical as it arrives at the observer.

A geometrical consideration shows that the radius of the circular arc is

$$R = \frac{r^2 + x_1^2 - (x_1 - b)^2}{2r} \tag{9}$$

giving

$$\sin\theta = \frac{2rx_1}{r^2 + x_1^2 - (x_1 - b)^2} \tag{10}$$

Let $\theta_0$ be the corresponding angle without the gravitational field, i.e.

$$\sin\theta_0 = \frac{r}{b} \tag{11}$$

Then eq.(11) may be written

$$\sin\theta = \frac{\sin\theta_0}{1 - \frac{b}{2x_1}\cos^2\theta_0} = \frac{\sin\theta_0}{1 - \frac{1}{2}b\hat{a}\cos^2\theta_0} \tag{12}$$

where $\hat{a} = 1/x_1$ is the acceleration of gravity at the position of the observer. Hence, as observed from above the sphere appears enlarged.

## 4b. A sphere seen from below

In this case the observer, P, is a distance $b$ vertically below the centre of the sphere at a position $x = x_1$. Figure 3 shows the situation in an arbitrarily chosen vertical plane through the centre of the sphere. Again we have drawn a circular arc representing a light ray emitted from the sphere and passing through the observer, making an angle $\theta$ with the vertical as it arrives at P, which is equal to the half of the apex angle.

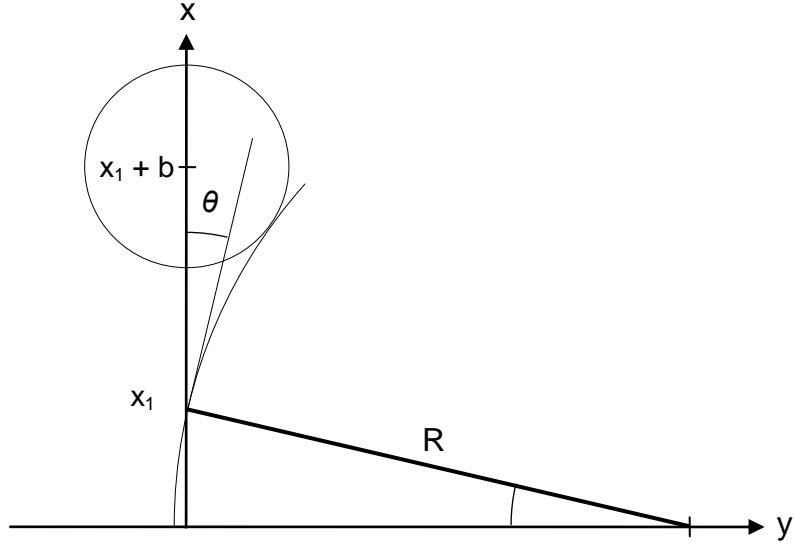

**Figure 3.** The figure shows the situation in an arbitrarily chosen vertical plane through the centre of the sphere. The circular arc is a light ray emitted from the sphere and passing through the observer, making an angle $\theta$ with the vertical as it arrives at the observer.

This time the radius of the circular arc is

$$R = \frac{\left(x_1+b\right)^2 - r^2 - x_1^2}{2r}, \qquad (13)$$

giving

$$\sin\theta = \frac{x_1}{R} = \frac{2r x_1}{\left(x_1+b\right)^2 - r^2 - x_1^2}. \qquad (14)$$

As in the previous case the corresponding angle without the gravitational field is given by eq.(11). Hence, eq.(14) may be expressed as

$$\sin\theta = \frac{\sin\theta_0}{1 + \dfrac{b}{2x_1}\cos^2\theta_0} = \frac{\sin\theta_0}{1 + \dfrac{1}{2}b\hat{a}\cos^2\theta_0}, \qquad (15)$$

Where $\hat{a}$ is the acceleration of gravity at the position of the observer. Hence, as observed from above the sphere appears diminished.

It may be noted that the photographed colour will vary from the centre of the sphere to the periphery. In the case of a static sphere and observer there will only be a gravitational frequency shift, depending on the height difference between the emission point and the observer. In the case 4a with the sphere beneath the observer there is a red shift. In this case the picture may show a sphere that appears yellow at the centre and red far from the centre. In the case 4b with the sphere above the observer there is a gravitational blue shift, and the same sphere may appear more bluish far from the centre.

## 5. A freely falling sphere in the Rindler space

We shall here reconsider the situation described by Beig and Heintzle[1]. They considered an observer and a sphere initially at rest in an inertial frame. The sphere emits light. At a certain point of time the observer starts accelerating, and moves subsequently with constant rest acceleration. Beig and Heintzle describes what the observers sees, especially how the observed angular extension changes, by taking aberration into account. In particular they find the somewhat surprising result that initially the angular extension of the sphere *increases* in spite of the fact that the observers moves away from the sphere. Their description is given with reference to the inertial frame in which the sphere is at rest.

In this section we shall consider the same situation. But we shall consider it from the point of view of the accelerated observer, and find out how he explains the observed phenomena.

In the uniformly accelerated frame R the observer and a sphere with radius $r$ is falling freely along the $x$-axis, both initially moving upwards. At the point of time $t = 0$ the observer and the sphere is at their uppermost position in R. From this point of time and onwards the observer is at rest in R at the position $x_1, 0, 0$, while the sphere proceeds falling freely. For $t > 0$ the motion of the center of the sphere is given by

$$x = \frac{h}{\cosh gt} \tag{16}$$

where $h = x_1 - b > 0$. Here $b$ is the distance between the observer and the center of the sphere when the sphere is instantaneously at its uppermost position.

As described in the inertial frame IF the situation is as follows: The sphere is at rest with the center at the position $X = h$, $Y = Z = 0$. For $T > 0$ the observer is moving hyperbolically in the positive $X$ – direction starting at the point $X = x_1$ at the point of time $T = 0$. Hence the position of the observer is given by

$$X^2 - T^2 = x_1^2 \tag{17}$$

or

$$\tilde{a} X = \left(1 + \tilde{a}^2 T^2\right)^{1/2} \tag{18}$$

where $\tilde{a} = 1/x_1$ is the proper acceleration of the observer. Introducing the proper time $\tau$ of the observer, we have

$$\tilde{a} X = \cosh \tilde{a} \tau , \quad \tilde{a} T = \sinh \tilde{a} \tau \tag{19}$$

We now imagine that the observer in R receives light at the point of time $t_1$ coming tangentially from the surface of the falling sphere. The event that the light is received is called event (1) and has coordinates $(t_1, x_1, 0, 0)$. The emission of the light from the surface of the sphere is called event (2) and has coordinates $(t_2, x_2, y_2, z_2)$. The even (2) is found by transforming to IF and then backwards to R. In IF the light moves along a straight path tangent to the sphere which is at rest. In IF the event (1) is given by

$$T_1 = x_1 \sinh g t_1 , \quad X_1 = x_1 \cosh g t_1 , \quad Y_1 = Z_1 = 0 \tag{20}$$

The angle $\vartheta$ is the half of the apex-angle as measured by a new observer which imagined to be at rest at point (1) in Figure 4. Hence,

$$\sin \vartheta = \frac{r}{X_1 - h} = \frac{r}{x_1 \cosh g t_1 - h} \tag{21}$$

For event (2) we find

$$T_2 = T_1 - r \cot \vartheta = x_1 \sinh g t_1 - r \cot \vartheta \tag{22a}$$

$$X_2 = h + r \sin \vartheta , \quad Y_2 = r \cos \vartheta , \quad Z_2 = 0 \tag{22b}$$

where $\vartheta$ is given in eq.(21). From this we get by means of the transformation (1),

$$x_2 = \left[ X_2^2 - T_2^2 \right]^{1/2} = \left[ \left(h + r \sin \vartheta\right)^2 - \left(x_1 \sinh g t_1 - r \cot \vartheta\right)^2 \right]^{1/2} ,$$
$$y_2 = r \cos \vartheta , \quad z_2 = 0 \tag{23}$$

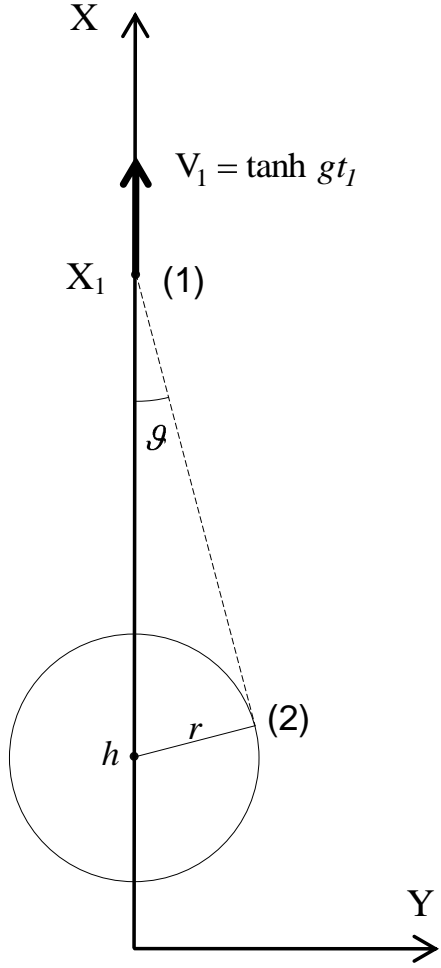

**Figure 4.** The apex-angle of a sphere beneath an observer at rest at a position $X = X_1$ in IF.

We now know two points, $x_1, 0, 0$ and $x_2, y_2, 0$ on the circular path of the light. The center of the circular arc is at the $y$-axis, and is found from simple geometry. Its position is (see fig. 5)

$$y_0 = -\frac{x_1^2 - x_2^2 - y_2^2}{2y_2} \quad , \quad \tan\theta = -\frac{x_1}{y_0} \tag{24}$$

We shall now show that the angle $\theta$ is equal to the corresponding angle $\theta'$ in IF which Beig and Heintzle[1] found by using the aberration formula. The angle $\vartheta$ in eq.(21) is the half apex angle for an observer at rest in IF. Taking aberration into account we find for the half apex angle $\theta'$,

$$\sin\theta' = \sqrt{1-V_1^2}\,\frac{\sin\vartheta}{1-V_1\cos\vartheta} = \frac{\sin\vartheta}{\cosh gt_1 - \sinh gt_1 \cos\vartheta} \tag{25a}$$

$$\tan\theta' = \frac{\sin\vartheta}{\cosh gt_1 \cos\vartheta - \sinh gt_1} \qquad (25b)$$

Inserting the expressions (23) into eq.(24) we find that $\tan\theta$ in eq.(24) is equal to $\tan\theta'$ in eq.(25b), hence that $\theta = \theta'$. This means that the apex angle found in the inertial reference system IF by taking aberration into account is equal to the apex angle found in the accelerated Rindler frame by taking into account that light moves along circular paths in this system due to the acceleration of gravity experienced in the Rindler space.

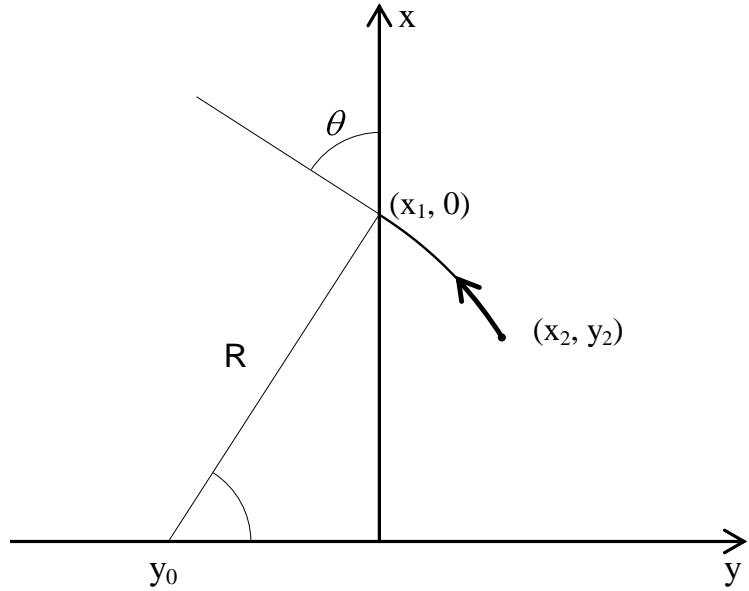

**Figure 5.** Relationships used to deduce eq.(24).

The optical phenomena described by Beig and Heintzle[1] associated with a uniformly accelerated observer, have been explained from the point of view of an inertial observer by these authors. We have here provided the description of the same phenomena as perceived by the accelerated observer. According to the general principle of relativity the observer may consider himself as at rest. And according to the principle of equivalence the (Newtonian) inertial acceleration field he experiences has the same physical effects as a field of gravity due to a mass distribution. Hence he explains the change of the apex-angle of the sphere as due to a combination of its motion away from him and the action of the field of gravity upon the motion of light, which causes the light paths to be spherical instead of straight, as we are used to.

It should be noted, however, that there is a restriction upon this description due to the horizon in the Rindler frame. In eq.(16) we must demand that $h > 0$, which requires that $x_1 > b$ or $\tilde{a}b < 1$. This condition is also found in the treatment of Beig and Heintzle in the form $ax_0 < 1$. According to Beig and Heintzle this is the condition that the apex-angle as a function of the proper time of the observer shall have a maximum. Such a maximum also exists in our treatment where the observer is at rest and the sphere is falling. In R the existence of this maximum has the following simple explanation.

Initially the observer and the sphere move freely upwards with a fixed proper distance between them. At the point of time $t = 0$ both reaches a maximal height. The observer remains at rest in R for $t > 0$, and the sphere proceeds falling downwards. The light received by the observer at $t = 0$ was emitted by the sphere a little earlier, when the sphere had not yet arrived at its uppermost position. Hence the apex-angle increases until the emission time is $t_2 = 0$. When light emitted at this point of time arrives at the observer at a later point of time $t_1$ he will see a maximal apex-angle.

Let us consider this in some detail. Inserting $T_2 = 0$ into eq.(22a) we get

$$x_1 \sinh gt_1 = r \cot \vartheta \tag{26}$$

where $\cot \vartheta$ is expressed in terms of $t_1$ in eq.(21). We arrive at the following equation for $t_1$,

$$x_1 \sinh gt_1 = \left[ x_1 \cosh gt_1 - h^2 - r^2 \right]^{1/2} \tag{27}$$

The solution is

$$\cosh gt_1 = 1 + \frac{x_1 - h^2 - r^2}{2x_1 h} \tag{28}$$

Here $h = x_1 - b$. From eq.(2) follows that the proper time of the observer is

$$\tau = gx_1 t = gt / \tilde{a} \tag{29}$$

Hence

$$g t_1 = \tilde{a} \tau_1 \tag{30}$$

Utilizing the formula $\sinh \alpha / 2 = \left[ \cosh \alpha - 1 / 2 \right]^{1/2}$ we find the proper time $\tau_1$ at the time that he observes a maximal apex-angle, is given by

$$\sinh \frac{\tilde{a}\tau_1}{2} = \frac{\tilde{a}}{2}\sqrt{\frac{b^2 - r^2}{1 - \tilde{a}b}} \qquad (31)$$

This is in agreement with the point of time of the observation of a maximal apex-angle deduced by Beig and Heintzle[1] with reference to IF. It is interesting that the reason for such a maximum is totally different in IF and in R, and that its existence is more obvious, as explained above, in R than in IF, where it is rather surprising.

Beig and Heintzle[1] consider one more strange phenomenon: the finite limit of the apex-angle for an observer in infinite hyperbolic motion. In our description the motion of the sphere is given by eq.(16), and the observer is at rest in R above the sphere at the height $x = x_1$. That is, the sphere is at rest in IF and the observer is in hyperbolic motion away from the sphere with proper acceleration $\tilde{a} = 1/x_1$.

According to eq.(16) the sphere approaches the horizon of the Rindler frame at $x = 0$ in the limit $t \to \infty$. The velocity

$$\frac{dx}{dt} = -bg\frac{\tanh gt}{\cosh gt} \qquad (32)$$

is zero in this limit. Thus, for $t \to \infty$ we have the situation shown in Figure 6.
A simple geometrical consideration shows that

$$R = \frac{x_1^2 + r^2}{2r} \qquad (33)$$

and

$$\sin \theta_\infty = \frac{x_1}{R} = \frac{2\tilde{a}r}{1 + \tilde{a}^2 r^2} \qquad (34)$$

Here $\theta_\infty$ is the limiting apex-angle. This is the same result as the one found by Beig and Heintzle[1] in their eqs.(20') and (26b).

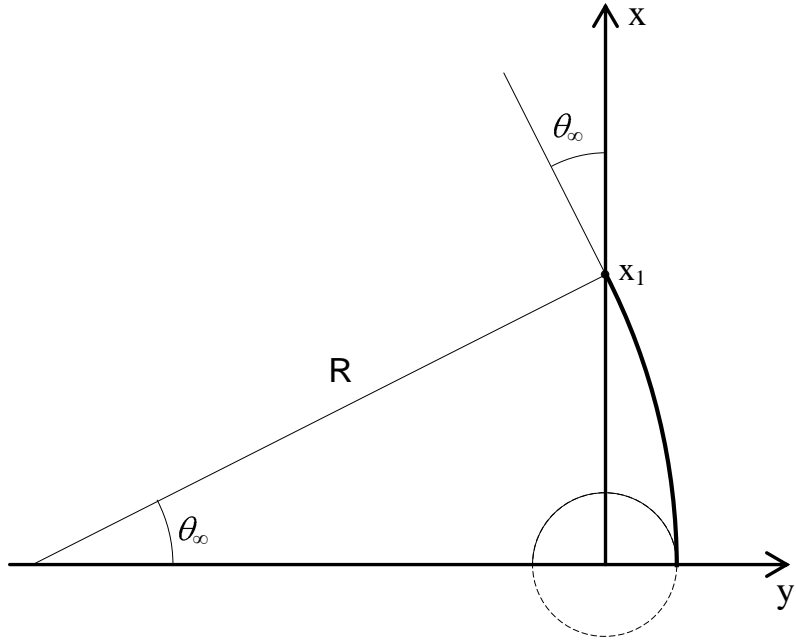

**Figure 6.** The situation defining the limiting angle $\theta_\infty$.

Again the reason for an observable result is different in IF and in R. In IF the reason that the limiting angle does not vanish when the sphere is infinitely far from the observer is that its velocity approaches that of light and the aberration effect conspires with the vanishing geometrical extension of the sphere to produce a finite limiting apex-angle. In R the reason that the apex-angle is finite in the limit $t \to \infty$ is that the sphere never comes farther down than to the horizon. Note, however, that both in IF and in IF the sphere will become invisible in the far away future because of an infinite redshift. In IF this is due to the Doppler-effect, and in R it is due to the gravitational frequency shift.

## 6. Conclusion

In this article we consider optics in a gravitational field. In the simple case of a parallel gravitational field as experienced in a uniformly accelerated reference frame this is an optics where straight light rays are replaced by circular rays of light.

The present work was inspired by a recent article by Beig and Heintzle[1]. They deduced some surprising optical phenomena associated with the visual appearance of a sphere at rest in an inertial frame as seen by a uniformly accelerated observer. The phenomena were explained with reference to the inertial frame of the sphere by taking aberration into account.

In this article we have described the observable phenomena with reference to the co-moving Rindler frame of the accelerated observer, and we have found how the observer would explain the phenomena.

Due to the principle of equivalence our analysis also provides insight as to the optics in a permanent gravitational field due to a mass distribution. In general the laws of geometrical optics connected for example with lenses and mirrors have to be modified since light does not follow straight lines, but circular paths. However the radius is of the arcs is great. In a uniform field with acceleration of gravity $g = 10 m/s^2$ the radius of the paths is approximately $c^2/g = 1\, l.y.$. Hence it will not be simple to perform an experiment showing that light does not follow straight lines in a gravitational field. But in principle it can be done.

**References**

1. R. Beig and J. M. Heinzle, *Relativistic Aberration for Accelerated Observers.* Am. J. Phys., **76**, 663-670 (2008).
2. E. Eriksen and Ø. Grøn, *Electrodynamics of hyperbolically accelerated charges V: The field of a charge in the Rindler space and the Milne space.* Ann. Phys. **313**, 147-196 (2004).

**Acknowledgement**

We would like to thank Mari Mehlen for help with the figures.